\documentclass[page-classic]{epl2}

\title{Effective surface-tension in the noise-reduced voter model}

\author{Luca Dall'Asta\inst{1,2} \and Claudio Castellano\inst{3}}
\shortauthor{Luca Dall'Asta \etal}

\institute{
 \inst{1} Laboratoire de Physique Th\'eorique (UMR du CNRS 8627),
    B\^atiment 210, Universit\'e de Paris-Sud, 91405 ORSAY Cedex (France) \\
 \inst{2} Abdus Salam International Center for Theoretical Physics, 
    Strada Costiera 11, 34014, Trieste (Italy) \\
 \inst{3} Dipartimento di Fisica, Universit\`a ``La Sapienza'' and SMC-INFM,
    P.le A. Moro 2, 00185 Roma, (Italy) \\
}


\abstract
{
The role of memory is crucial in determining the properties
of many dynamical processes in statistical physics.
We show that the simple addition of memory, in the form of noise reduction,
modifies the overall scaling behavior of the voter model, introducing an
effective surface tension analogous to that recently observed in memory-based
models of social dynamics.
The numerical results for low-dimensional lattices show a scaling
behavior in good agreement with usual Cahn-Allen curvature-driven coarsening, 
even though slower preasymptotic regimes may be observed depending on the
memory properties.
Simple arguments and a mean-field analysis provide an explanation for
the observed behavior that clarifies the origin of surface tension and
the mechanism underlying the coarsening process.
}

\begin{document}

\maketitle

The theory of phase-ordering kinetics \cite{bray} describes the scaling
behavior of domain coarsening phenomena occurring in non-equilibrium
statistical physics, with applications ranging from social sciences
\cite{opinion} to ecology~\cite{tilman}.
The way in which a system orders starting from a disordered phase follows
different paths depending on generic properties (e.g. symmetries,
the type of interactions, etc.) that allow to identify specific universality
classes of nonequilibrium phase-ordering dynamics. 
In particular, the important family of `kinetic Ising models'
with $Z_{2}$-symmetry and no strict conservation of the order parameter
displays two different ordering behaviors, depending on whether a surface
tension exists or not. If this is the case (as in Glauber dynamics, for example),
coarsening is curvature-driven~\cite{dornic}:
the domain length scale grows following a power law, $\ell(t) \sim t^{1/2}$
in any dimension~\cite{bray}.
Other models, such as the voter model~\cite{liggett}, exhibit domain
coarsening without surface tension: dynamics is driven by interfacial
noise~\cite{dornic}.
In this case $\ell(t)$ grows as $t^{1/2}$ in one dimension,
logarithmically in $d=2$ and it does not diverge for $d>2$~\cite{krapivsky}.
It is also possible to define a family of kinetic Ising models interpolating
between zero-temperature Glauber dynamics (T0GD)
and the voter model (VM) ~\cite{dornic,drouffe,mendes}:
in these models, the transition rates depend on two parameters measuring
the strength of bulk and interfacial noise. When bulk noise is present
the coarsening is curvature-driven. When only interfacial noise exists,
coarsening is voter-like for a broad class of models,
either characterized by global conservation of magnetization or
by nonconserving $Z_2$-symmetric dynamical rules~\cite{dornic}.

In this Letter, we introduce a simple model,
consisting of a voter-like dynamics modified by a noise-reduction procedure. 
The main motivation of this work consists in understanding the mechanism
governing  the ordering process of some recently proposed models of social
dynamics,  such as the Naming Game \cite{baronka,baronka2} and the
Wang-Minett model of bilingualism \cite{minett,sanmiguel}.
In both these models, the pairwise interaction rules are reminiscent of the
voter dynamics, but the update follows a two-steps process
involving memory effects.
Surprinsingly, on $d$-dimensional lattices, these models undergo
a domain coarsening dynamics following approximately the same
temporal laws of curvature-driven models with non-conserved order parameter.
In order to elucidate the origin of this effective curvature-driven
coarsening, we define a minimal model, that we call ``noise-reduced
voter model'' (NRVM), in which a noise-reduction prescription is added
to the usual interaction rule of the VM, introducing a memory effect.
As it will be clear below, the introduction of a noise-reduction prescription
does not give rise to bulk noise in the spin dynamics. 
Instead, it turns out that the additional ingredient gives rise to an effective,
memory-induced, surface tension, that generates in low dimension
a curvature-driven coarsening process.

Noise reduction has been used in the study of surface
growth~\cite{eden} and diffusion-limited aggregation~\cite{DLA} as a method
to speed up the approach to asymptotic scaling.
It amounts to attaching to each site of the system a counter
recording how many times a particle has touched the site.
Only when the counter reaches a given threshold value the site becomes
active and the particle sticks to it.
We introduce the noise reduction in the framework of the VM in the
following way:
each site $i$ is endowed with a spin variable, $s_i$, as well as with a pair of
counters $C_i^+$ and $C_i^-$. At each time step, a site $i$ and one of its
neighbors $j$ are selected at random to interact. The counter associated
with the spin value of the selected neighbor is increased by one:
$C_i^{s_j} \to C_i^{s_j}+1$. At odds with the normal VM dynamics,
the spin variable $s_i$ is not necessarily modified upon interaction.
It is updated only when one of the counters attains an a priori
fixed threshold value $r$ (e.g. if the positive counter reaches the
value $C_i^+=r$ then $s_i$ is set equal to $+1$). 
In this way the interaction between spins is indirect, being mediated by
the dynamics of counters.
Two possible update rules for the counters are considered: when 
the spin at site $i$ flips, only the counter that
has reached $r$ is set to zero (model 1), or instead both counters
are reset (model 2).

For $r=1$ both models coincide with the usual voter dynamics, whose behavior
is well known in any dimension~\cite{krapivsky} and is characterized by the
absence of surface-tension~\cite{dornic}.
For $r>1$, one expects an effective surface-tension to arise: the repeated
interaction with neighbors,
needed before a spin update is actually performed, implies that the counters
sample the local field of each site. According to this
argument, in the long run, local field imbalances are accumulated,
pushing the system evolution rule towards the T0GD. 
In fact, the NRVM does not simply interpolate
between the VM and the T0GD: the transition rates
encode the memory on how the spin pattern has been reached, thus it is
easy to find examples of configurations such that the rates
are not intermediate between regular voter and Glauber dynamics. 

\begin{figure}
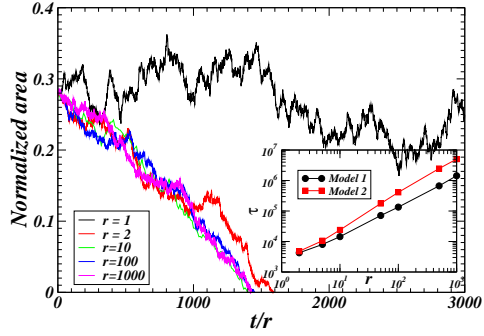

\onefigure[width=0.45\textwidth]{Fig1.eps}

\caption{Illustration of the presence of surface tension in
the two-dimensional voter model with noise reduction.
The initial condition is a configuration ($L=500$) with a droplet
of radius $R_{0}=200$ surrounded by a sea
of opposite spins.
We have monitored the normalized area $\pi R^2(t)/L^2$ as a function of time
for model 1.
This quantity decreases linearly for $r>1$, as expected in the presence of
surface-tension~\cite{bray}.
Note that the linear trend is clear even if the data refer to a single
realization.
Time is rescaled by $r$ to show that the noise-reduction parameter simply
rescales the temporal evolution.
In the inset, the average time $\tau$ needed to dissolve the droplet is plotted
as a function of $r$ for both models: $\tau$ scales linearly for model 1
and with an effective exponent 1.1 for model 2.}
\label{fig1}
\end{figure}

In order to ground more firmly the intuition about the existence
of a surface tension, we have checked the presence of surface tension in a
two-dimensional lattice model by studying the evolution of a circular droplet
in a large system of $N=L^2$ sites. 
Figure~\ref{fig1} shows that, as for systems with curvature-driven
coarsening~\cite{bray}, the volume of the droplet decreases linearly
with time, i.e. the typical size of a domain decreases as $\sqrt{t}$.
The figure refers to model $1$ dynamics, but a similar behavior is observed
even for model $2$.
Moreover, rescaling time with $r$ the curves for
different values of $r$ are very close, revealing that changing
the threshold value produces only a linear temporal
rescaling in the ordering process.\\
An explanation of the linear rescaling can be argued from the
analysis of the equation for the dynamics of counters.
The probability that a spin $s_{ i}$ flips is given by the joint
probability that a neighboring site of opposite spin is chosen and
the counter $C_{i}^{-s_{i}}=r-1$. On a $d$-dimensional lattice, the
probability to choose a neighbor of $i$ with opposite spin value is $W(s_{i})$
=  $\frac{1}{2}\left[ 1- \frac{1}{2 d} s_{i}
\sum_{j \in \mathcal{V}(i)} s_{j} \right]$,
where $\mathcal{V}(i)$ is the neighborhood of $i$. 
In model 1, the dynamics of the two counters at a site $i$ are decoupled, thus the probability
$p^{\pm s_{i}}_{n}(t)$ that the counter $C_{i}^{\pm s_{i}}$ is equal to a value
$n$ at time $t$ is defined by the equation,
\begin{equation} \label{evolP2D}
\frac{\partial}{\partial t} 
p^{\pm s_{i}}_{n}(t) = - \left[p^{\pm s_{i}}_{n}(t) - p^{\pm s_{i}}_{n-1}(t) \right]
W(\mp s_{i})~,
\end{equation} 
with periodic boundary conditions on the variable $n$.
The solution of this equation requires the knowledge of the probability
$W(s_i)$ at all times, i.e. of the whole dynamics but, formally solving the equation in the Fourier space, 
we can write
\begin{equation}\label{sol2D}
p^{\pm s_{i}}_{r-1}(t) = \frac{1}{r} \left\{ 1+ \sum_{k=1}^{r-1} \cos\left(2\pi k/r+ \sin(2\pi k/r) \mathcal{W}^{\mp}_{i}(t) \right) e^{- [1-\cos(2 \pi k/r)] \mathcal{W}^{\mp}_{i}(t)} \right\},
\end{equation} 
where $\mathcal{W}_{i}^{\pm}(t) = \int_{0}^{t} W(\pm s_{i}(t')) dt'$.
Note that the sum in the r.h.s. of Eq.~\ref{sol2D} tends to 0 when the 
integral is much larger than $r$, while it tends to -1 when it is much smaller.
For a spin in the bulk of a domain $W(-s_{i})=1$, while $W(s_i)=0$.
Hence, in the former case the integral $\mathcal{W}_{i}^{-}(t)$ diverges and $p^{s_i}_{r-1}(t) \to 1/r$,
while in the latter the integral $\mathcal{W}_{i}^{+}(t)$ is finite, so that $p^{-s_i}_{r-1}(t) \to 0$.
As a consequence the effect of varying $r \gg 1$ can be reabsorbed in
a temporal rescaling $t \to t/r$ (see the inset of Fig.~\ref{fig1}). 
In order to extend the analysis to model 2, one should consider the equation for 
the joint probability $q^{\pm s_{i}}_{n_{+},n_{-}}(t)$ that the counters $C_{i}^{\pm s_{i}}$ assume some values $n_{+}$ and $n_{-}$, since the counters 
are now coupled. We will consider such approach later for the simpler case of a mean-field model with $r=2$.
However, we find numerically that in $d=2$ the scaling is not far from the one for model 1.
In $d=3$, changing $r$ in simulations apparently gives more
than a simple rescaling of time (see below), but this is probably
just a preasymptotic effect.

\begin{figure*}
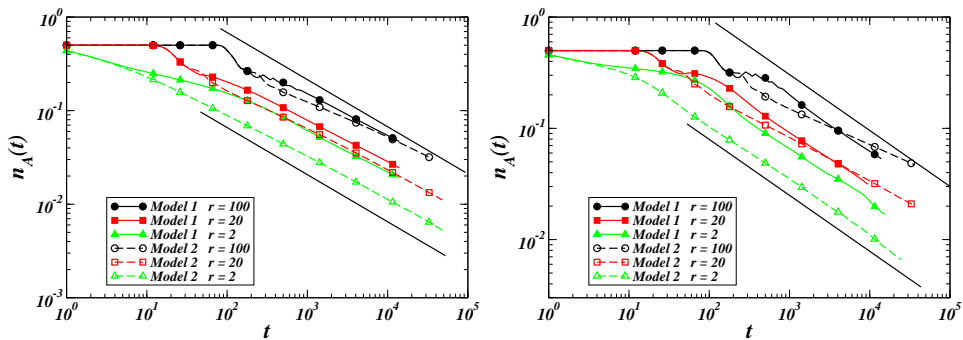

\centerline{
\includegraphics[width=0.45\textwidth]{Fig2a.eps} 
\includegraphics[width=0.45\textwidth]{Fig2b.eps}
}
\caption{Temporal evolution of the density of interfaces $n_{A}(t)$ in the
voter model with noise reduction on a two-dimensional lattice of
size $L=5000$ (left) and on a three-dimensional lattice of size $L=300$
(right).
The straight lines, reported for reference, have the slope $1/2$ expected for
curvature-driven coarsening.}
\label{fig2}
\end{figure*}

Another remark arises considering the master equation for the probability
distribution $P(S,t)$ of having a certain $d$-dimensional configuration 
$S=\{s_{i}\}$ at the time $t$ in model 1, 
\begin{equation}
\label{distr2D} 
\frac{d}{d t} P(S,t) = \sum_{i} \left[ W(-s_{i})
P(S^{i},t) p^{s_{i}}_{r-1}(t) - W(s_{i}) P(S,t) p^{-s_{i}}_{r-1}(t)\right],
\end{equation}
where $P(S^{i},t)$ is the probability of the configuration $S$ with the
only spin $s_{i}$ flipped. A similar equation holds for model 2. 
From Eq.~\ref{distr2D}, we can write the evolution equation for the
average quantities, such as the local magnetization $\langle s_{i} \rangle$
\begin{equation}\label{evolS2D}
\frac{d \langle s_{i}\rangle}{dt} \propto \langle p^{-s_{i}}_{r-1}(t)
\nabla^2_{i} s_{i} \rangle~, 
\end{equation}
in which $p^{-s_{i}}_{r-1}(t)$ evolves according to Eq.~\ref{evolP2D}.
Eq.~\ref{evolS2D} is similar to the analogue equation for the usual voter
model~\cite{krapivsky}, but the presence of the counter probability $p^{-s_{i}}$
breaks down the conservation of the average magnetization, which is
one of the crucial features of the voter model on regular lattices.
 
\begin{figure*}
\centerline{
\includegraphics[width=0.45\textwidth]{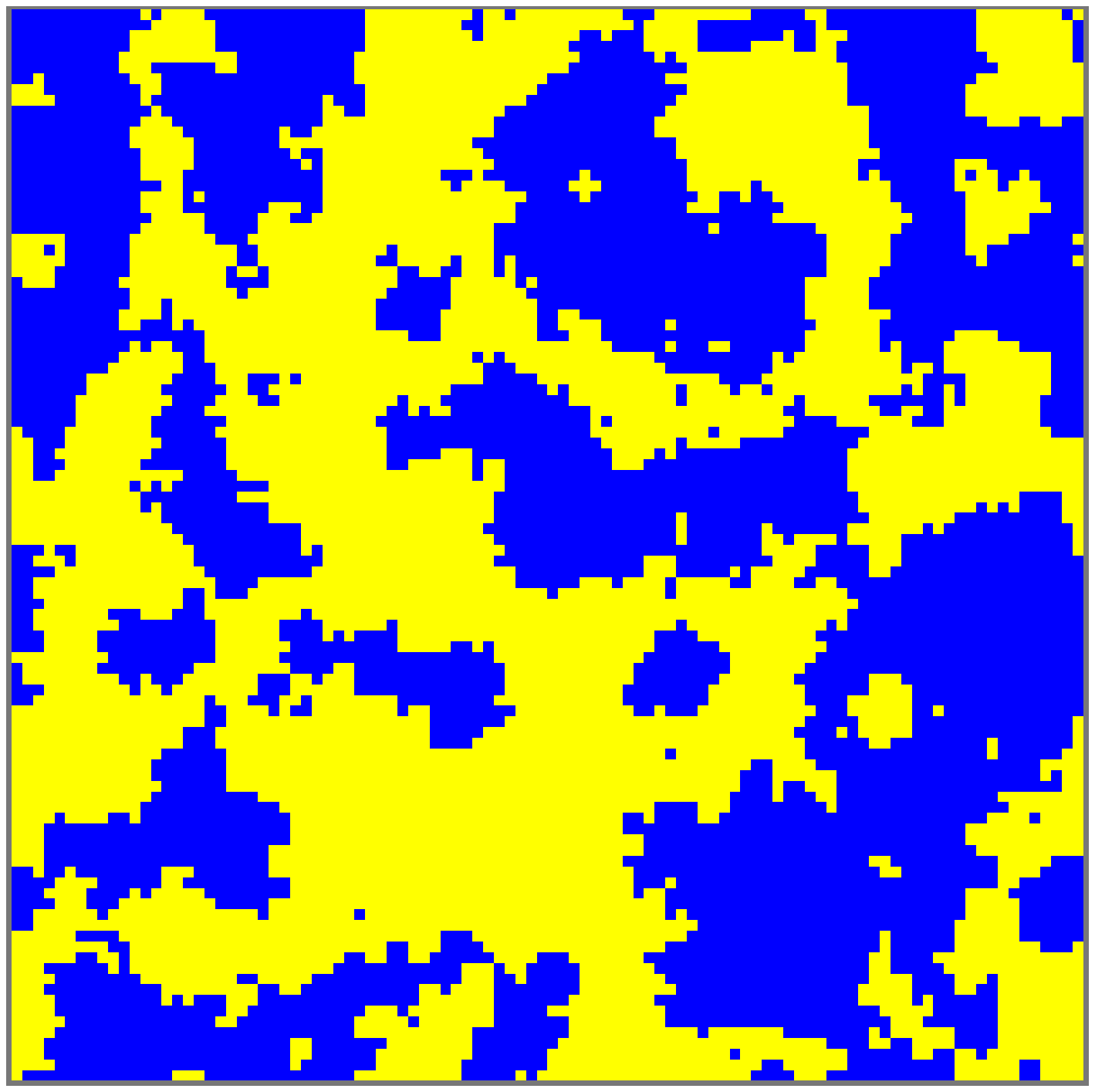} ~~~
\includegraphics[width=0.45\textwidth]{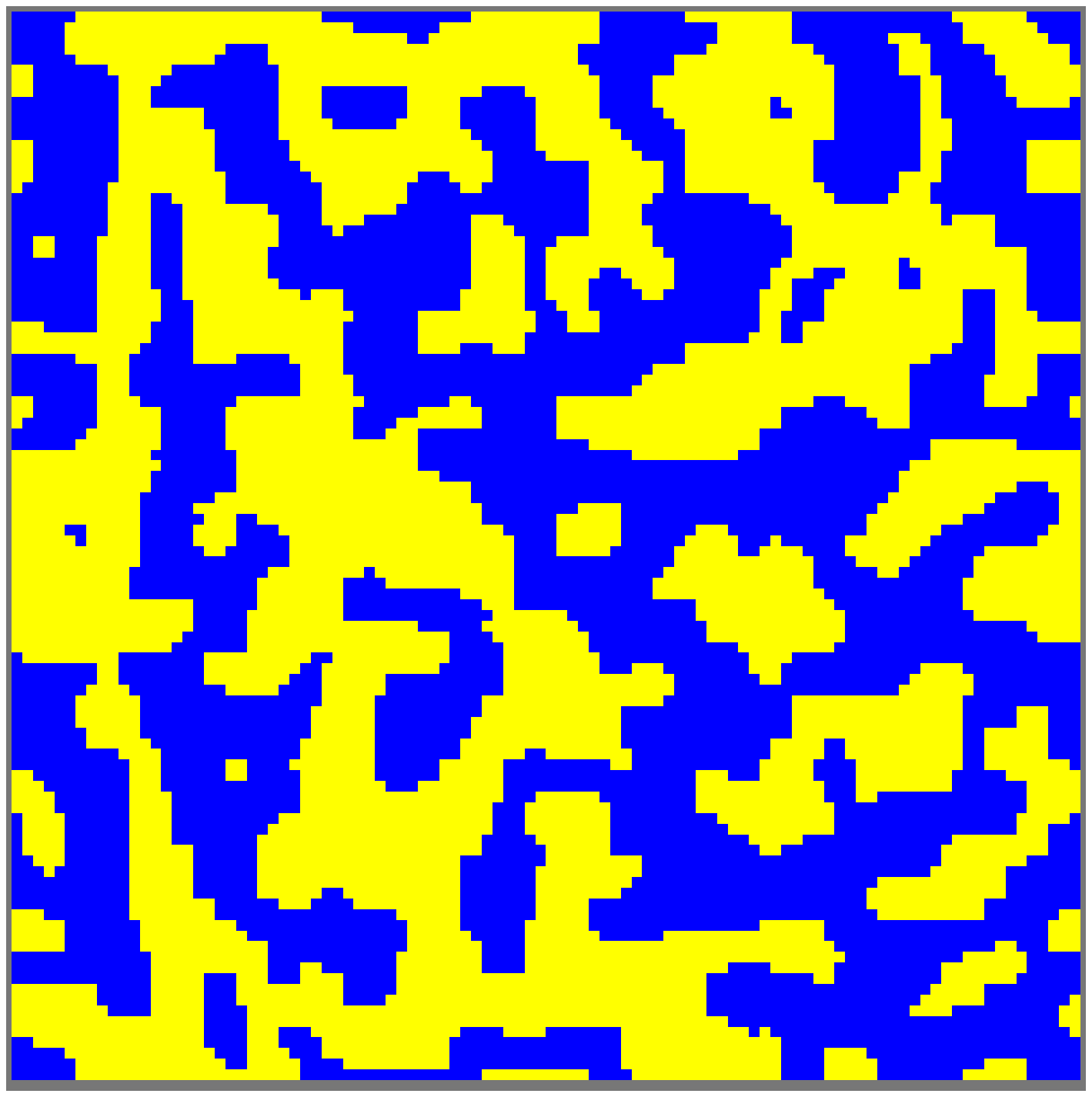} 
}
\caption{Snapshot of a spin configuration for model 2, $r=2$ (left)
and $r=1000$ (right) showing that domain interfaces tend to be faceted
for large $r$. The two configurations have comparable total length
of the interface.
}
\label{Pattern}
\end{figure*}

The presence of an effective surface tension is expected to induce a
curvature-driven coarsening, resulting in a phase-ordering process with
scaling properties belonging to the Cahn-Allen universality class~\cite{bray}.
This has been checked directly by studying numerically the coarsening dynamics
in $d=2$ and $d=3$ starting from a fully disordered random configuration with
zero magnetization.
The temporal evolution of the density of interfaces, $n_A(t)$, reported in
Fig.~\ref{fig2}, shows in general a power-law decay
$n_{A}(t) \propto t^{-1/z}$ for all cases.
For model 1, after the trivial initial transient, the effective exponent
sets to about 0.45 in $d=2$ and close to the value 1/2 typical of the
Cahn-Allen universality class in $d=3$.
Remarkably, the value 0.45 coincides with the value found recently
by Castell\`o et al.~\cite{sanmiguel} in another modified voter model.
We cannot ascertain numerically whether the discrepancy with 1/2 is
the consequence of logarithmic corrections or a truly asymptotic exponent.
For model 2, the violation of the curvature-driven decay (exponent 1/2)
is much stronger. In $d=3$ a numerical fit gives an effective
exponent that decreases with $r$: 0.39 for $r=20$ and 0.33 for $r=100$.
The origin of this change can be understood by visual inspection
of the pattern of coarsening domains (Fig.~\ref{Pattern}).
For large $r$ the interfaces between the two phases tend to be faceted
up to a length $\ell_c(r)$, that grows as $r$ is increased.
At the beginning of the ordering process, when the typical size of
domains $\ell(t)$ is smaller than $\ell_c(r)$, the coarsening process
is strongly perturbed (and slowed down) by the faceted nature of
interfaces. This explains the reduced value of the effective exponent.
Later on, when the scale of domains becomes sufficiently large
($\ell(t) \gg \ell_c(r)$) the short scale structure of domain boundaries
does not matter anymore, interfaces will be rough and therefore we expect
curvature-driven coarsening (with its associated exponent 1/2) to rule.
This picture applies well to the two-dimensional case.
The effective exponents found for large $r$ in $d=3$ are probably due
to an extremely long crossover to the asymptotic regime, although
present numerical evidence is not clear-cut in this respect.

To understand the dynamical interplay between the voter dynamics
and the nonequilibrium surface-tension, it is useful to perform a
mean-field analysis, considering the NRVM on the complete graph.
Following Ref.~\cite{Lavicka03}, we derive a master equation
for the probability $P(m,t)$ that the magnetization takes the value
$m$ at time $t$. In a mean-field framework, single-site counters are well 
represented by the average values $C^{\pm}$ over the whole system.
We call $p^{\pm}_{n}(t)$ the mean-field probabilities that the
average counter value $C^{\pm}$ is equal to $n$ at time $t$.
In a temporal step, the magnetization can change of a value $\pm 2/N$ or
remain constant.
The corresponding probabilities appearing in the master equation are,
\begin{equation}\label{prob1}
Prob\left\{ m \to m \pm \frac{2}{N}\right\}=
\frac{1+m}{2}\frac{1-m}{2} p^{\pm}_{r-1}(t).
\end{equation}
Writing down the master equation, 
using Eq.~(\ref{prob1}), expanding the expressions up to the second
order in $1/N$ (with $N \to \infty$), and neglecting higher order terms, we get
\begin{eqnarray} \label{masterEQ}
\nonumber \frac{\partial}{\partial t} P(m,t) = &
 - & \frac{1}{2N} \left[p^{+}_{r-1}(t)-p^{-}_{r-1}(t)\right]
   \frac{\partial}{\partial m} \left[ (1-m^2) P(m,t) \right]\\
& +& \frac{1}{2N^2} \left[p^{+}_{r-1}(t)+p^{-}_{r-1}(t)\right] 
 \frac{\partial^2}{{\partial m}^2} \left[ (1-m^2) P(m,t) \right]~.
\end{eqnarray}
Two other equations for the counter probabilities $p^{\pm}_{n}(t)$
are required to close the equation for $P(m,t)$. For model 1, the equations have the same form of Eq.~\ref{evolP2D},
while in the case of model 2 one should start from the evolution equation for the mean-field
joint probability $q_{n_{+},n_{-}}(t)$, noting that $p^{\pm}_{r-1}(t) = \sum_{n_{\pm}} q_{n_{+},n_{-}}(t)$.
The solution of the coupled set of equations is, in general, beyond reach, but
it is instructive to look at Eq.~\ref{masterEQ}:
apart from the counters dependent prefactors,
the second term is the one present in the equation for the
VM~\cite{Lavicka03},  while the first is instead the one found for
T0GD~\cite{Castellano05}, expressing
the existence of a drift, due to the imbalance between the counters.
The interplay of the two terms reflects the competition between
curvature-driven and noise-driven coarsening.
\begin{figure}
\centerline{
\includegraphics*[width=0.5\textwidth]{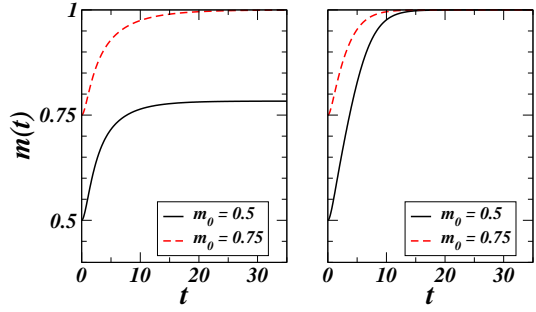} 
}
\caption{Temporal evolution of the magnetization $m(t)$ for model 1 (left)
and model 2 (right) on a complete graph, using rate equations.
Two curves for different initial magnetization $m_{0}$ reach different
stationary states in model 1.
}
\label{fig3}
\end{figure}

The rate equations for the mean-field magnetization $m(t)$ can be similarly derived;
let us elucidate the simple cases for $r=2$. In model 1, the following system of rate equations
holds
\begin{eqnarray}\label{ratem0}
\dot{m}(t) &=& \frac{1-{m(t)}^{2}}{2} \left[p^{+}_{1}(t) - p^{-}_{1}(t)\right]~,\\
\label{ratep} \dot{p}^{\pm}_{1}(t) &=& \frac{1\pm m(t)}{2} \left[ 1 - 2 p^{\pm}_{1}(t)\right] ~.
\end{eqnarray}
The case of model 2 is more complicate, since we need to compute the four coupled evolution
equations for $q_{n_{+},n_{-}}(t)$.
The corresponding system of rate equations reads
\begin{eqnarray}\label{ratem1}
\dot{m}(t) &=& \frac{1-{m(t)}^{2}}{2} \left[ q_{1,0}(t) - q_{0,1}(t)\right]~,\\
\dot{q}_{0,0}(t) &=& - q_{0,0}(t) + \frac{1+m(t)}{2} q_{1,0}(t) + \frac{1-m(t)}{2} q_{0,1}(t) + q_{1,1}(t)~,\\
\dot{q}_{1,0}(t) &=& - q_{1,0}(t) + \frac{1+m(t)}{2} q_{0,0}(t) ~,\\
\dot{q}_{0,1}(t) &=& - q_{0,1}(t) + \frac{1-m(t)}{2} q_{0,0}(t) ~,\\
\dot{q}_{1,1}(t) &=& - q_{1,1}(t) + \frac{1+m(t)}{2} q_{0,1}(t) + \frac{1-m(t)}{2} q_{1,0}(t) ~,
\end{eqnarray}

Linear stability analysis indicates
that the two models present different stationary solutions.
For model 2, there are only two possible stable stationary solutions,
$\{m=-1, q_{0,0}=1/2, q_{1,0}=0, q_{0,1}=1/2,
q_{1,1}=0\}$,
and  $\{m=+1, q_{0,0}=1/2, q_{1,0}=1/2, q_{0,1}=0,
 q_{1,1}=0\}$,
i.e. the system completely orders into a ferromagnetic phase. The solution with $m=0$ is instead unstable.
Hence, in the model with double counter update, the mean-field dynamics
is completely governed by the drift term.
On the contrary, in the model 1, fluctuations play a relevant role.
The possible stationary states are
$p_{\pm}=1/2$ for all possible values of $m$,
and $m=1$ and $p^{+}_{1}=1/2$ for any
value of $p^{-}_{1} \leq 1/2$
(or the symmetric case
$m=-1$, $p^{-}_{1}=1/2$, $p^{+}_{1} \leq 1/2$).
The existence of multiple stable stationary states points to the existence
of a transition depending on the initial value of the magnetization $m(0)$.
Indeed, solving numerically Eqs.~(\ref{ratem0}-\ref{ratep})
we find 
\begin{equation}\label{mtransition}
\overline{m} =
\left\{\begin{array}{cc} \pm m(0)/m^*  & \mbox{for    } |m(0)| < m^* \\
                                        \pm 1 & \mbox{for    } |m(0)| \geq m^*
\end{array} \right.
\end{equation}
while the difference $p^{+}_{1}-p^{-}_{1}$ is 0 below the transition
and finite for $|m(0)|>m^*$ (Fig.~\ref{fig3}).
The transition point is $m^*\approx 0.6382\ldots$.\\
If one performs simulations on the complete graph, the sharp transition
is actually smeared out by fluctuations. 
Nevertheless we can observe a rather abrupt crossover between a regime where
$p^{+}_{1}-p^{-}_{1}$ is exponentially small
(for $|m(0)| \le 0.22\ldots$)
to a regime with finite $p^{+}_{1}-p^{-}_{1}$, whereas
asymptotically $|m|$ is equal to 1 in both cases.
The shift in the position of this transition is due to the diffusive
contribution of the second term 
in l.h.s. of Eq.~\ref{masterEQ}, that is neglected in the rate equations.
For small $|m(0)|$ the fully ordered state is not reached because
the initial condition is not sufficiently asymmetric and the
drift, which decreases during the evolution (both $p^{+}_{1}$ and $p^{-}_{1}$
tend asymptotically to $1/2$), is not strong enough to fully order
the system.
If $|m(0)|$ is large enough instead, the counter imbalance is sufficient
to drive the system to full order, keeping an asymptotic nonvanishing drift.
For $r>2$ the general phenomenology remains similar, with the
difference that the counters probabilities may exhibit oscillations
that in turn induce a nonmonotonic behavior of the magnetization.

In conclusion, we have introduced a noise-reduced voter model,
in which memory effects induce an effective surface tension. 
While in finite dimension this implies that the coarsening process is always
curvature-driven, on a fully connected graph it is possible to observe a
crossover between voter-like and zero temperature Glauber behavior. 
We have provided an explanation of the mechanism generating surface
tension, and numerical results supporting the thesis that the NRVM ordering
process 
belongs to the Cahn-Allen universality class of coarsening with non-conserved
order parameter~\cite{bray}. Further investigations on model 2 are required 
in order to clarify the asymptotic value of the coarsening exponent for
large $r$.
The present analysis is consistent with recent results obtained by 
Castell\'o et al.~\cite{sanmiguel}, showing that the existence of 
an intermediate state (here due to the introduction of memory) is relevant 
in modifying the scaling properties of the VM. 
In order to give further insight on more realistic situations, such as those
involving memory-driven non-equilibrium ordering dynamics on social
networks, it would be interesting to study the behavior of the model on
systems with heterogeneous topologies. We expect that the competition of 
the two types of kinetic mechanisms, i.e. curvature-driven coarsening 
and interfacial noise, may give rise to nontrivial ordering phenomena.

\end{document}